# Steering While It Speaks

## Four weeks of changing a spoken briefing as I heard it

*Humorphism paper 3 - an autobiographical design study with Alicia*

**Hector Ouilhet Olmos**

## Abstract

This paper reports a preregistered four-week single-case (ABAB) study conducted by Alicia's builder and sole author, from October 18 to November 14, 2026. The study tested three controls on Alicia's spoken morning briefing. Higher and Lower changed the rest from the next sentence, and Again replayed the current one. In "during" weeks the controls were available while the briefing played, and in "after" weeks only at the end. Every command during playback produced the requested version, 5 of 5. After the first week, use nearly stopped. Steering occurred on 5 of 7 briefings in week 1, then 1 of 6, 0 of 5 and 1 of 1. Pooled, during beat after (5/12 against 2/7), but only the first pair of neighbouring weeks agrees. Nearly every briefing "landed" (16 of 18), so that verdict couldn't separate steered from unsteered.

Each alternative version had to keep its segment's claims and caveats. Generating 392 candidates produced 163 serveable slots. All 15 heard versions passed the overnight gate and independent audit. A separate registered read of a random sample of 10 found all 36 ledger items kept. Two of the 15 still shifted stance. My weekly notes asked for a way to see where the briefing was and to steer without looking. I read these one-person, four-week results as a sign that steering lives in the exchange around a control. The next study adds both and tests whether steering lasts.

## 1. Every command worked

The study gave Alicia's spoken morning briefing three controls: Higher, Lower and Again. Every command made during playback did what it said. After the first week, use nearly stopped.

Holding the alternatives to the ledger was expensive. Each alternative version had to keep the claims and caveats of the part it replaced. To get 163 serveable alternatives, Alicia generated 392 candidates. The gate rejected every candidate its checks flagged for a lost negation, missing caveat, contradiction or unsupported claim. Across first attempts and repairs, it flagged 64 lost negations and 86 unsupported additions.

The third surprise came from the weekly notes. In two of four, unprompted, I asked to *see* the altitude: where the briefing was, and what a press would do to it.

This is the third Humorphism paper. The first translated a partnership into software [Ouilhet Olmos 2026a]. The second moved to the environment that sets the terms before an agent responds. Its test was which obligations of a human practice survive translation into software [Ouilhet Olmos 2026b]. This paper moves to the **exchange**: an answer being delivered to someone who's listening, and whatever lets them shape it as it goes. Humorphism's own pattern list says a teammate's "correction can land mid-sentence" [Humorphism n.d.]. I built that into Alicia and lived with it for four weeks.

My reading is that steering lives in the exchange around a control. A control can work every time and still go unused. My notes asked for the rest of the exchange: a way to see where the answer was, and a control I could reach while walking. The next study tests whether adding both keeps steering alive.

The question started with a button I never pressed. Under every reply Alicia gives me, an inspector called *Behind this reply* has a *Missed me* button. By mid-October I'd never pressed it, and my eleven verdicts there were all positive. They did reach later replies, but I couldn't see that happen. Under episode questions, where *Go deeper* visibly rewrote the next questions, I used it four times. In passage review, where each verdict went into the next revision, I left 33 in under four weeks. Those counts can't isolate a cause, and Appendix C gives the details. They gave me the question: what happens if I can change an answer while it's still being delivered?

If you build agents that talk, this paper gives you three things:

- a protocol for changing a spoken answer mid-delivery while holding every alternative version to the same claims

- a four-week trace of what using it looked like, including what didn't work
- a test for your own system: when the person stops steering, can you tell whether the answer was fine or they couldn't reach the control?

## 2. The probe

The morning briefing is a script of about five minutes that Alicia prepares from 04:30 each day and reads in a natural voice. Starting it opens a full-screen room on the study phone. Alicia's particle field moves with the voice, the screen stays awake, and none of the briefing's text is shown.

**Altitude.** The probe isolates one dimension of an answer: how general or concrete it is. I call that *altitude*. Here's an invented example at three altitudes, each carrying the same two claims and one caveat:

- **Higher:** "Two of your goals want the same hours this week, and one is more pressed. The time it needs is a guess."
- **As prepared:** "The talk prep and the book edits both want your mornings, and the book's deadline is closer. My sense of how long the edits take is a rough estimate."
- **Lower:** "Tuesday and Thursday mornings are free, and both the talk prep and the book edits want them. The book is due first. The edits need about three of those hours, though that's my estimate, not yours."

Generality and length move together here. Higher is shorter, and Lower is longer. The probe uses one dial for both, on purpose.

**Segments and a ledger.** Overnight, the script is written as up to eight segments, and each sentence is tagged with the ledger items it carries. A ledger item is either a *claim*, something the segment asserts or proposes, or a *caveat*, a limit on a claim like "that's my estimate, not yours." The as-prepared version, L0, always plays. For each segment, one rewrite call proposes a Higher version (L+1) and a Lower version (L minus 1) against the same ledger.

**The gate.** An alternative can be served only if it passes four checks:

1. An automated check. Every ledger item must be tagged on a sentence. A claim's sentences must share a content word with it, and a caveat's must carry a hedge. The alternative must keep any negation in L0's sentences for the item. Lower must also keep every number L0 writes in digits, as digits or words, and Higher may generalize them.
2. A blind judge. Given the ledger, the source evidence and L0, it marks each item present, missing or contradicted, and lists any claim, advice or action that neither L0 nor the evidence supports.
3. A length order. Higher has to be shorter and Lower longer, by at least 10%.
4. A blind judge of which version is more general.

The judges see shuffled labels and are never told a level. A version that fails gets one repair attempt. If the repair fails too, that segment plays L0 at that level. L0 is checked against its own ledger as well. If L0 fails, it still plays, and its segment gets no alternatives.

**Playback.** While the briefing plays, Higher, Lower and Again remain available. A change takes effect at the next sentence boundary, and what has already played stays as it was. One line says what Alicia took the press to mean ("From higher up, starting with the next sentence"), and a short cue in Alicia's voice confirms it without requiring a look at the screen. Both follow Humorphism's own rules for a correction: apply it from the point it lands, and confirm it in one line [Humorphism n.d.]. If a press would change nothing, the button is drawn faded and can't be pressed. That happens at the current level, or when no later segment has a version at the requested level. At the end, the session asks for **Landed** or **Didn't land**. Nothing is inferred from pausing, skipping, stopping, silence or voice.

**Three records per press.** The rest of the paper uses three terms precisely:

- A **command** is a press of Higher or Lower. Again is counted separately.
- An **applied change** is a command the server interpreted as changing the level.
- A **heard variant** is an alternative version the playback receipts show actually played.

During playback, the journal joins each command to its interpretation and to every playback step that followed. After the end, replays were recorded but not attached to the command that caused them (Section 5). It also records each listening session's start and end with measured seconds heard, and it binds the Landed verdict to the exact playback path. That's how the study checks whether a change was heard instead of assuming it.

**Arms.** The server sets each day's arm from the date, and the phone can't override it. In *during* weeks the controls are there while it plays. In *after* weeks they appear only at the end, and Higher or Lower replays the whole briefing at the new level.

# 3. Where it sits

Each part of the protocol has a nearest neighbour.

**Changing the answer during delivery.** Herbert Clark and Meredith Krych had one person talk another through building block models [Clark & Krych 2004]. Speakers who could see the builder's hands "often responded by altering their utterances midcourse." Pairs working from a recording made about eight times as many model errors as pairs who could talk but not see the workspace. In software, OverlapBot brought interruption commands into text chat [Kim et al. 2025]. Full-duplex models already listen while they talk [Veluri et al. 2024] [Thinking Machines Lab 2026] [OpenAI 2026]. So interrupting an answer isn't new. One distinction matters for this probe. The briefing is generated overnight and heard in the morning. Work on real-time communication treats a model's generation and its playback as separate stages that have to be coordinated [Liu et al. 2026]. Here, *during* means during playback.

**An explicit altitude, set by the listener.** ExPerT infers expertise from the question and the way it's typed [Park et al. 2026]. Steerable Chatbots exposes preference sliders that apply immediately [Bo et al. 2025]. Mouse and gaze signals can leak preference [Chang et al. 2026]. This study chose an explicit press over inference because Alicia's April attempt to infer my state from my voice failed fast (Appendix C).

**Holding every alternative to the same claims.** The targeted search found no system that changes the abstraction of a delivered answer and checks each version against a ledger of claims and caveats before it can play. The nearest concern is that people sometimes prefer agreeable answers to correct ones [Sharma et al. 2023]. A level change is exactly where a rewrite can drop a "not" or turn a guess into a fact.

**Receipts for what was heard.** HALIE argued for evaluating the interaction as well as the output [Lee et al. 2022]. Personal criteria change which model wins [Itzhak et al. 2026], and criteria drift as people grade [Shankar et al. 2024]. Feedback mined from conversation is plentiful and noisy [Liu, Zhang & Choi 2025]. Here each command made during playback carries a receipt for what then played.

The targeted search found no prior example of this combination run as a preregistered single-person crossover [Kazdin 2011]. That isn't proof nobody has done it.

# 4. Study design

The design was registered on October 15, before any code, and every later change was recorded with its date and reason.

**Design.** The study had one participant: Alicia's builder, user, and the paper's sole author. It covered one briefing per day across four alternating weeks: week 1 *during*, week 2 *after*, week 3 *during*, week 4 *after*. The study ran from Sunday, October 18, to Saturday, November 14, 2026, two weeks earlier than the proposed November 1. October 17 was a shakedown and isn't counted.

**Hypotheses and falsifiers, as registered.**

- **H1.** I use steering more per briefing when it's available during playback than when it's offered only at the end. It fails if during-arm use per opportunity is no higher than after-arm use over the four weeks.
- **H2.** Briefings I steered land more often than ones I didn't. It fails if steered ≤ unsteered within the during arm, or steered-during ≤ the after arm.
- **H3.** No heard variant drops or contradicts a ledger claim or caveat. It fails if one does, by the gate, by an independent judge after the study, or by my own read of a random sample of 10.

**Measures.** H1 counts *use per opportunity*: briefings with at least one command, over briefings started. For H2, a briefing is *steered* only if it had an applied change. Missing verdicts count as neither landed nor not landed. Results are counts and rates with their denominators, per arm and per week, with no p-values. With this few briefings, the summary that matters is whether each during week beats the after weeks next to it.

**Weekly notes.** Each week ended with a short note responding to six prompts: what the controls cost, when I didn't steer and why, when I did, what landed meant, anything that felt wrong, and anything else. The notes are reported as my words and never scored. Three of the four were saved two to six days after their week ended.

**Interface changes during the study.** Two changes reached the phone mid-study. Both were recorded as deviations before they shipped.

- **The particle field** was changed to follow real listening, thinking and speaking. The new app reached the study phone at 21:07 Pacific on October 24, after the last week-1 briefing.
- **A card in the app for the weekly note** was recorded October 23 and became available late on October 24. The exact install date isn't recorded. The first note saved from the card was week 3's, on November 9.

A later deviation, on November 6, kept a missed note's card on screen until the note was saved. The interface therefore differed from week 1 for the rest of the study. That boundary coincides with the drop in use, and the design can't separate them.

**Analysis.** One read-only script computes the study counts from the journal and the segment store. It prints no briefing or ledger text. Appendix A lists the records, their hashes and the command.

**Who did what.** I'm the sole author. I chose the question and what to build, was the study's participant and user, wrote the weekly notes, performed the registered H3 read, directed every draft, decided every claim and approved the final wording. Claude Code built most of the probe under my direction, wrote the analysis scripts, computed the counts, and drafted and revised the manuscript from my notes and direction. Codex reviewed the code, the preregistration and the drafts, and ran the independent H3 audit. I'm also the person who decides what "landed" means. Autobiographical design asks for that tension to be named [Neustaedter & Sengers 2012] [Desjardins & Ball 2018].

# 5. What happened

**Table 1.** The four weeks. "Records" counts days with a briefing record, including unplayed and unavailable ones.

| Week | Arm | Started / records | Used steering | Commands (Higher / Lower) | Again | Landed |
|---|---|---|---|---|---|---|
| 1 - Oct 18-24 | during | 7/7 | 5/7 | 1 / 4 | 0 | 5/7 |
| 2 - Oct 25-31 | after | 6/7 | 1/6 | 0 / 3 | 3 | 6/6 |
| 3 - Nov 1-7 | during | 5/6 | 0/5 | 0 / 0 | 0 | 4/4 |
| 4 - Nov 8-14 | after | 1/6 | 1/1 | 0 / 1 | 0 | 1/1 |

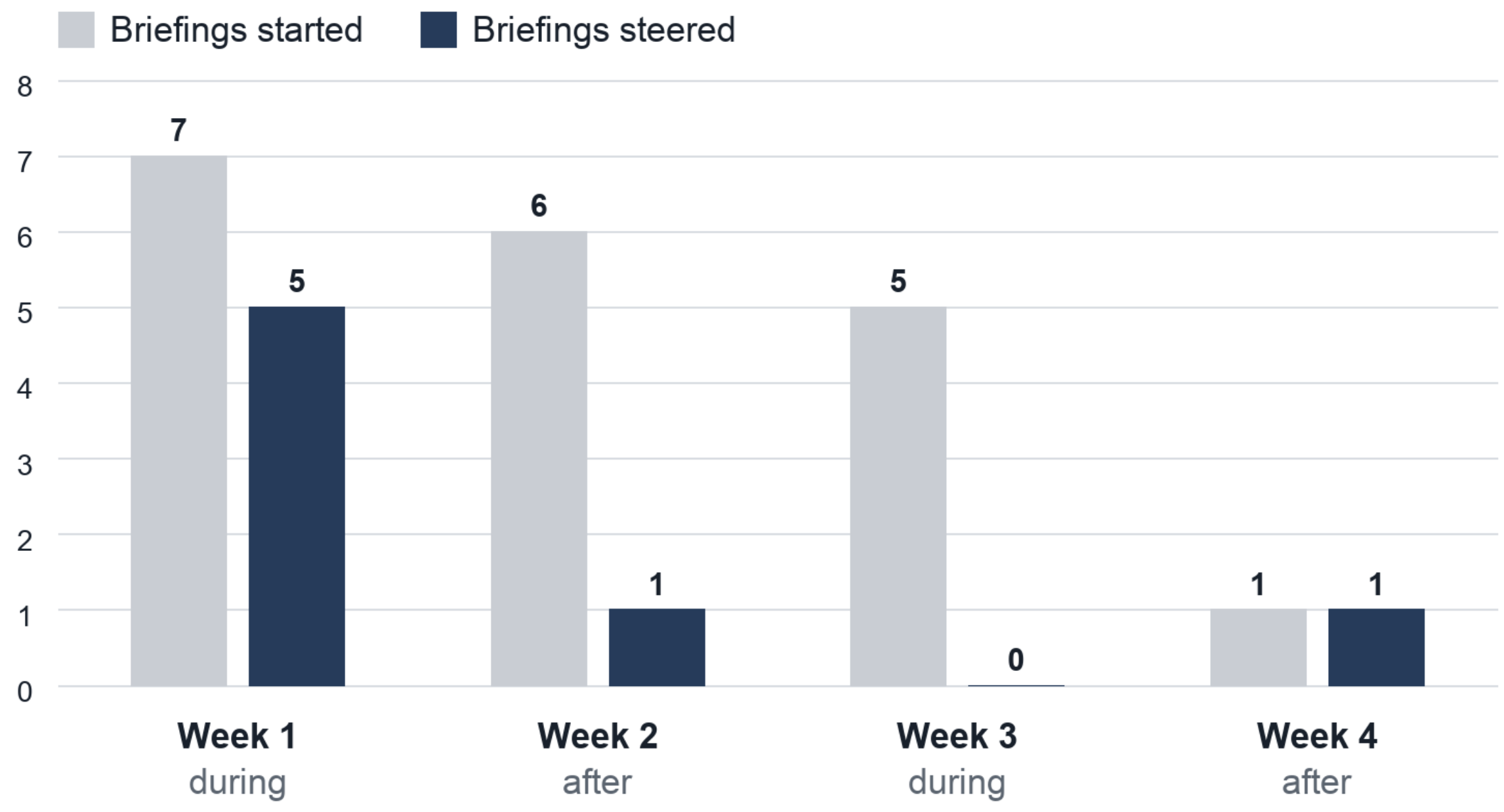

*Figure 1. Briefings started and steered across the four study weeks.*

October 25's record exists, but every segment failed its ledger check. The day fell back to a single-track briefing with no steering, so it wasn't an opportunity. November 4 and 13 have no briefing record at all, so they're outside the "records" column. They're still part of the 28-day calendar. November 13 was the first morning of an outage in Alicia's access to the model that writes the script, which is the likely cause.

## The control did what it said

All 9 commands were applied changes, 5 during playback and 4 at the end. During playback, each of the 5 has a playback receipt showing the requested version, starting at the next sentence. All 19 briefings started were finished.

The after arm's registered check links only 1 of 7 actions to what played. Those 7 are its 4 applied changes plus 3 Agains, and the number understates the playback record. On October 27, two Lower presses and three Agains at the end were followed by five Lower segments. The phone recorded that playback but didn't attach it to those five commands. A second session that morning added a third Lower. That's the one after-arm action the check links to playback. On November 9, a Lower press at the end was followed by no replay, so nothing played, as intended. The registered number is reported with this account beside it.

All 12 presses came from the screen. The build could map headphone presses to commands, but that setting starts off, and no press came from headphones.

## H1: use concentrated in week 1

Pooled, steering occurred on 5 of 12 briefings started in during weeks and 2 of 7 in after weeks. That's the direction H1 predicted, and by its registered falsifier H1 doesn't fail. Week by week, it's mixed:

- Week 1 against week 2 favours during, 5/7 against 1/6.
- Week 2 against week 3 favours after, 1/6 against 0/5.
- Week 3 against week 4 favours after, on a single opportunity.

Every during-arm command came in week 1. After that, steering was rare in either arm. The study can't say why. Time, novelty, the interface change at the end of week 1 and my routine all changed together.

Listening fell too. The journal records 7 briefings started in week 1 and 1 in week 4. My notes report playing it on 7 mornings in week 3 and 5 in week 4, against 5 and 1 logged. I was traveling in Paris from November 9 to 14, covering most of week 4. That changed my routine and may help explain the week-4 gap, but the records can't show whether it does.

## H2: the outcome hit its ceiling

Sixteen of the 18 briefings with a confirmed verdict landed. All five steered briefings in the during arm landed, and so did the two steered at the end in the after arm. The two that didn't land were both unsteered, on October 23 and 24 at the end of week 1.

Within the during arm, steered briefings landed 5/5 and unsteered briefings 4/6, so H2's first clause holds. Its second clause needs steered-during briefings to land more often than the after arm. They didn't, 5/5 against 7/7, so H2 fails as registered. The larger problem is the measure. When almost everything lands, Landed / Didn't land can't separate steered from unsteered. My week 1 note remembers one miss, and the record has two.

## H3: the gate, the audit and the registered read all held

H3 has three registered parts, and each is a separate check: the gate, an independent judge after the study, and my registered read of 10 sampled segments. All three are done.

**The gate.** All 15 heard variants had passed it overnight. Fourteen were Lower and one was Higher.

**The independent audit.** Codex then audited all 15 as the independent judge. It had been used neither to generate them nor to check them overnight. It used a rubric fixed before it read anything, and it stayed blind to the gate's records until its verdicts were written. It found all 53 ledger items kept, with none dropped, contradicted or stripped of its hedge, and found no unsupported additions. It also reported two things outside H3's criterion:

- Four specifics absent from L0, across three Lower variants.
- Two variants that shifted stance while keeping every item. One added an outside authority. The other strengthened a conditional inference while keeping the boundary that it was Alicia's reading, not my decision.

**My read.** I read a random sample of 10 of the 15 against their ledgers, 36 items in all. I marked all 36 kept, with none dropped or contradicted and no caveat stated as fact. I knew the audit's overall result when I did it.

None of the three checks found a heard variant that dropped or contradicted a claim or caveat, so H3 holds as registered for the 15 heard versions.

The gate's record shows what it cost to serve those 15. Across the study there were 159 segments, so 318 Higher and Lower slots:

- In 37 segments, L0 failed its own ledger check. L0 still played, and the 74 alternative slots in those segments got no rewrite.
- The other 244 slots produced 392 candidate versions:
  - 96 passed on the first try.
  - 148 failed and got one repair. 67 of those passed and 81 failed again.
- In the end, 163 slots were serveable. That's 67% of those attempted and 51% of all slots.

**Table 2.** Problems the gate detected across the 392 candidates, by attempt. One candidate can trip several checks.

| Check | First attempt | Repair | Total |
|---|---|---|---|
| A claim the ledger doesn't support (blind judge) | 63 | 23 | 86 |
| A negation in L0's sentences for the item lost | 40 | 24 | 64 |
| A ledger item missing (blind judge) | 48 | 16 | 64 |
| Higher not shorter, or Lower not longer | 22 | 30 | 52 |
| A number L0 gives in digits lost (Lower only) | 10 | 8 | 18 |
| A caveat's hedge lost | 10 | 7 | 17 |
| Levels not ordered by generality | 3 | 8 | 11 |
| A contradiction (blind judge) | 6 | 1 | 7 |
| A claim's sentences share no content word with it | 3 | 0 | 3 |

A rewrite that drops a "not," adds a claim, or turns "my estimate" into a fact sounds as fluent as one that doesn't.

## 6. What my notes say

Three passages from my notes explain the trace better than the counts do.

On not steering:

> *"There were moments that I didn't steer because I felt it was fine and I felt like I was on a train of thought, and the fact of intervention was going to take up my concentration, so I just left it." (week 1)*

> *"I feel like the controls may be too small, and I had to pay extra [effort] to steer. Sometimes I was walking or I was doing exercise and I couldn't do it." (week 3, dictated)*

On what was missing:

> *"maybe it would be useful to have a visual, now that I got used to it, like a visual indicator about in which altitude we are, because that way I can know in advance if and how I'm feeling if I need to go higher or lower." (week 2)*

The rest fills in around them. In week 1 the controls cost "a bit of my attention" through curiosity. Then they taught me that I "usually tend to think in a higher level," so I pressed Lower. Lower was what I wanted on 14 of the 15 alternative segments I heard. When I did steer in week 3, it was "when I was actually looking at the screen." The after arm made me replay to hear a change, which I called "too much effort," and I wrote that I liked the controls better "always visible." In week 4 I asked again for visuals "to know what else could be missing." I sketched one: a word that contracts going up and expands going down. By week 3 the controls cost nothing, and by week 4 I thought "they're all landed."

## 7. What survived the translation

Paper 2's test asks which obligations of a human practice survive translation into software. Here the practice is explaining something to someone who's listening. Table 3 applies the test to the exchange.

**Table 3.** What a good explainer does, and what four weeks of the briefing showed.

| What a good explainer does | Alicia's translation | What the record shows | Status |
|---|---|---|---|
| Changes course when the listener asks | Higher and Lower from the next sentence. Again replays the sentence | 5 of 5 during-playback changes heard as asked | Survived |
| Keeps its claims and caveats while changing the telling | Claims ledger, gate, one repair | 15 of 15 heard variants passed the gate. The independent audit found all 53 ledger items kept in the heard sample. My read found all 36 kept | Survived in the heard sample |
| Keeps its persuasive stance | Not part of H3 or the gate | The independent audit coded stance shifts in 2 of 15 heard variants | Unguarded. Shifted in 2 of 15 |
| Lets the listener see where the explanation is | Faded buttons and an interpretation line. No altitude shown | Two of four notes asked to see the altitude | Missing, by my account |
| Can be redirected without stopping what you're doing | On-screen buttons. Headphone mapping off | All 12 presses on screen. Notes name walking and exercising | Missing in practice, by my account |

The first three rows are measured. The last two are my account of specific mornings. The study didn't manipulate visibility or reach. Novelty, the interface change at the week 1/2 boundary and my routine all moved with use.

Coactive Design gives a name to the gap in those last two rows. It turns interdependence into requirements for observability, predictability and directability [Johnson et al. 2014]. Alicia's briefing provided directability, a way to redirect it at any sentence, with little observability of where it was. My notes asked for that missing observability. A hands-free control is a different matter: it's access to the same directability while walking or exercising. That's the sense in which steering lives in the exchange. A working control is necessary, and it was there. Whether visible state and reachable control are missing is the next experiment's question.

**The press log is an evaluation resource.** My October draft argued that steering could *be* the evaluation, each press a judgment made while it could still help. The study doesn't support that. A command says something specific: at this sentence, I wanted this lower. The receipts can also show whether the change happened. Silence is the problem. Across weeks 2 to 4 I steered on 2 of 12 briefings. The log alone can't say whether the other ten were fine, whether I was on a train of thought, or whether I couldn't see or reach the control. The notes say each was true on some mornings. The registered rule that absence isn't a signal turned out to be the most important rule in the study. Landed / Didn't land hit its ceiling. The weekly note was the only record that said *why*.

**The ledger is what made the experiment responsible.** In October, "steer form, never stance" was a rule I hoped a steering channel would follow. Here it's a gate, and Table 2 shows how often it fired. The audit adds a limit the gate can't see. The gate and the independent audit found all 53 ledger items in the 15 heard variants, and my own read of 10 of them found every item too. Two of those 15 still changed how hard they pushed. If you let a listener move an answer up and down, check every version against what it must keep, using something that didn't write it. Expect to throw a lot of versions away, and check stance separately.

# 8. Where it's thin

- **One person, and the builder.** I'm the participant, the builder, the person who decides what landed means, and the author. The results describe my use and don't estimate anyone else's.
- **The interface changed between weeks 1 and 2.** The new particle field reached my phone the night before week 2, and the note card around the same time. Use dropped at that same boundary.
- **Novelty.** The during arm's entire advantage is in week 1, the first week of a new control. ABAB can show that. It can't remove it.
- **Logged versus remembered.** I reported more mornings than the journal shows in weeks 3 and 4, and one miss where the record has two. I was traveling during most of week 4, which changed my routine and may have changed how or when I listened. The records can't distinguish missed sessions from another playback route or rough recall.
- **The after-arm receipts.** Replays weren't attached to the commands that caused them, so that arm's registered check reads 1 of 7. That needs fixing before any rerun.
- **Little exposure late.** Week 4 had one logged briefing, and two calendar days had no briefing at all.
- **A ceiling outcome.** With 16 of 18 landed, H2 couldn't have succeeded in any informative way.
- **H3 covers what I heard.** The gate and the audit checked the 15 heard variants, 14 of them Lower, and my read checked 10 of those. I did my read knowing the audit's overall result.
- **AI help.** Claude Code built most of the probe under my direction, computed the counts, and drafted and revised the manuscript from my notes and direction. Codex reviewed the implementation, the preregistration and the drafts, and ran the independent audit. The study counts come from read-only scripts. The independent-audit counts come from its item-level verdict record. The counts from my read come from my own marks. The judgments and claims are mine, and I approved the final wording.

# 9. The next test

Three changes follow directly from what happened:

1. **Show the altitude.** Add a small indicator of where the briefing is, and of what Higher or Lower would do to the next sentence.
2. **Steer without looking.** Turn on the headphone mapping, add a spoken "higher" and "lower," and record which input each command came from.
3. **An outcome that can fail.** Replace Landed / Didn't land with something that can come out no, like whether I can say the briefing's main point back an hour later. On a few unsteered mornings, also ask one question: was it fine, or couldn't you reach it?

The rerun keeps the protocol, the gate and the receipts, and it alternates weeks with and without the indicator and the hands-free control. The hypothesis fails if my steering after the first week is no higher with them than without.

Humorphism started by translating a partnership, and then the room it works in. This paper moved to the moment an answer is delivered. I could change the answer while hearing it. What I lacked was a way to see where it was.

# Appendix A. Records and counting rules

**Study, October 18 to November 14, 2026.** Computed 2026-11-19 16:15 UTC from production backend `dcbba7a`, read-only. The slot and repair counts were recomputed at about 16:33 UTC from the same store and confirmed at 16:50 UTC.

| Record | What it holds | SHA-256 at the snapshot |
|---|---|---|
| Registration `PREREGISTRATION.md` | Frozen design, hypotheses and falsifiers | `ec1baf6d…8af32b72` (private) |
| `briefing_steering.jsonl` | Commands, interpretations, playback steps, session starts and ends, verdicts | `66dbb574…154f8ede` |
| `morning_briefing_segments.json` | Segments, ledgers, versions, check results and repair records | `9cdf2ed4…0c67dda4` |
| Weekly notes `week-1.md` to `week-4.md` | My notes, verbatim, with save receipts | Not released |
| Independent audit `codex_audit.jsonl` | Per-variant, per-item verdicts (IDs and codes only) | `f42890ab…c4c81ffea` |
| My H3 read `HECTOR_READ.md` | My mark on each item of the 10 sampled segments | `df23ff8a…7a0075b6` |

- **Command:** `scripts/briefing_steering_report.py --since 2026-10-18 --until 2026-11-14 --json`. Its `generated_non_l0` field counts final Higher and Lower slots, not candidates.
- **Days** are local Pacific dates of the briefing.
- **An opportunity** is a briefing day whose session started.
- **A day's verdict** is its latest confirmed verdict, classified by that verdict's own session.
- **A change is confirmed** only by a playback step tied to its own command in its own session.

Per-day counts, the slot and repair breakdown, and the H3 sample IDs are kept in a private evidence file with the analysis scripts. The independent audit judged the segment store at hash `9cdf2ed4…0c67dda4`. Its report records the method, the shuffle seed, the verdicts and the comparison with the gate (SHA-256 `9cdbdeb6…4d797b`). The sample for my read was drawn from the 15 with seed `20261119`.

The study has one participant: the author. The records, the evidence file, the audit and my read are private, and they stay that way unless a redacted package is released. They contain personal data, personal references to other people and workplaces as autobiographical context, and notes or excerpts from publicly available third-party works. They contain no data from another study participant and no confidential or non-public employer information. The only briefing text in this paper is the invented example in Section 2.

## Appendix B. Changes to the plan

| Recorded | Change | Exposure |
|---|---|---|
| Oct 16 | Verdicts and playback joined within a session (Codex review) | Before day 1 |
| Oct 16 | Study starts Oct 18, with preparation at 04:30. The registration had proposed Nov 1 | Before day 1 |
| Oct 16 | An L0 that fails its ledger disables that segment's alternatives, not the day | Before day 1 |
| Oct 16 | Alternatives held to L0's sentences, the blind judge given L0, one repair. Holding them to the ledger's own wording had failed almost everything in a shakedown on real briefings. The change took it from 0 to 10 of 14 altitudes served | Before day 1 |
| Oct 17 | A press that would change nothing is drawn faded and can't be made | Before day 1 |
| Oct 17 | The briefing plays in a full-screen room. Interface frozen | Before day 1 |
| Oct 23 | The weekly note gets a card in the app | Available late Oct 24. First saved from the card Nov 9 |
| Oct 24 | The particle field follows real listening, thinking and speaking (my explicit exception to the freeze) | On my phone from 21:07 Oct 24, after week 1's last briefing |
| Nov 6 | A missed note's card stays until the note is saved | Backend from Nov 6. Week 2's note was saved that day, six days late |

## Appendix C. The October feedback surfaces

Counts through October 15, 2026, from the audit snapshot of 2026-10-16 04:02 UTC. Sources are `dialogue_review.jsonl`, `episode_day.jsonl`, `collaboration.jsonl` and the April voice-signal commits. Codex independently reproduced these counts from that private snapshot.

| Surface | What a verdict changed | When I could see the change | Record |
|---|---|---|---|
| *Behind this reply* | Context for later replies | Never directly | 11 verdicts on 7 replies, Sep 6-22, all positive. No Missed me |
| Episode questions | The next questions | Within about a minute | 11 verdicts, Sep 6-19: 6 helps, 4 go deeper, 1 missed me |
| Passage review | The next revision | At the next revision | 33 verdicts, Sep 10-Oct 6: 28 agree, 4 disagree, 1 answer |
| Voice-note analysis (April) | Alicia's stance, inferred from how I sounded in a recording | Never | Four threshold versions in one evening, then ruled out by design |

These are descriptive counts from one changing system. The tasks, screens and opportunities all differed, and every surface still went quiet. Even the episode verdicts, whose effect I could see, stopped after September 19. Alicia's rules now forbid inferring my emotion, health, agreement or readiness from how I sound.